\documentclass[sigconf]{acmart}

\usepackage{booktabs} % For formal tables

\setcopyright{none}
\acmConference[GECCO '19]{the Genetic and Evolutionary Computation Conference 2019}{July 13--17, 2019}{Prague, Czech Republic}
\acmYear{2019}
\copyrightyear{2019}

\newcommand{\psl}{\texttt{PSL}}
\newcommand{\pamper}{\texttt{PaMpeR}}
\newcommand{\etal}{\textit{et al.}}
\newcommand{\induct}{\texttt{induct}}
\newcommand{\induction}{\texttt{induction}}

\begin{document}
\title{Towards Evolutionary Theorem Proving for Isabelle/HOL}
%\title{SIG Proceedings Paper in LaTeX Format}
%\titlenote{Produces the permission block, and
%  copyright information}
%\subtitle{Subtitle}
%\subtitlenote{The full version of the author's guide is available as
%  \texttt{acmart.pdf} document}
\author{Yutaka Nagashima}
\authornote{Supported by the European Regional Development Fund under the project AI \& Reasoning (reg. no.CZ.02.1.01/0.0/0.0/15\_003/0000466)}
\affiliation{%
  \institution{University of Innsbruck, Czech Technical University}
%  \city{Innsbruck}
%  \state{Tyrol}
%  \country{Austria}
}
%\affiliation{%
%  \institution{Czech Technical University}
%  \city{Prague} 
%  \country{Czechia} 
%}

%\renewcommand{\shortauthors}{Anonymous}

\begin{abstract}
Mechanized theorem proving is becoming the basis of reliable systems programming and rigorous mathematics. 
Despite decades of progress in proof automation, 
writing mechanized proofs still requires engineers' expertise and remains labor intensive. 
Recently, researchers have extracted heuristics of interactive proof development 
from existing large proof corpora using supervised learning. 
However, such existing proof corpora present only one way of proving conjectures, 
while there are often multiple equivalently effective ways to prove one conjecture. 
In this abstract, we identify challenges in discovering heuristics 
for automatic proof search and propose our novel approach to improve heuristics 
of automatic proof search in Isabelle/HOL using evolutionary computation.
\end{abstract}

%<ccs2012>
%<concept>
%<concept_id>10011007.10011074.10011784</concept_id>
%<concept_desc>Software and its engineering~Search-based software engineering</concept_desc>
%<concept_significance>500</concept_significance>
%</concept>
%<concept>
%<concept_id>10011007.10011074.10011099.10011692</concept_id>
%<concept_desc>Software and its engineering~Formal software verification</concept_desc>
%<concept_significance>300</concept_significance>
%</concept>
%</ccs2012>
%\end{CCSXML}



\maketitle

\section{Background}

\subsection{Interactive Theorem Proving}
Interactive theorem provers (ITPs) are forming the basis of reliable software engineering.
Klein \etal{} proved the correctness of the seL4 micro-kernel using Isabelle/HOL \cite{sel4}.
Leroy developed a verified opimizing C compiler, CompCert, in Coq \cite{compcert}.
Kumar \etal{} built a verified compiler for a functional programming language, CakeML, in HOL4 \cite{cakeml}.
In mathematics, mathematicians are substituting their pen-and-paper proofs 
with mechanized proofs to avoid human-errors in their proofs:
Hales \etal{} mechanically proved the Kepler conjecture using HOL-light and Isabelle/HOL \cite{kepler}, whereas
Gonthier \etal{} proved of the four colour theorem in Coq \cite{4colour}.
In theoretical computer science,
Paulson proved G{\"{o}}del's incompleteness theorems using Nominal Isabelle \cite{incomplete}.

\subsection{Meta-Tool Approach for Proof Automation}

To facilitate efficient proof developments in large scale verification projects,
modern ITPs are equipped with many sub-tools, such as proof methods and tactics.
For example, Isabelle/HOL comes with 160 proof methods defined in its standard library.
These sub-tools provide useful automation for interactive proof development.

\paragraph{\psl{}}
%Previously, 
Nagashima \etal{} presented \psl{},
a \underline{p}roof \underline{s}trategy \underline{l}anguage \cite{psl}, for Isabelle/HOL. 
\psl{} is a programmable, extensible, meta-tool based framework, 
which allows Isabelle users to encode abstract descriptions of how to attack proof obligations.

Given a \psl{} strategy and proof obligation,
\psl{}'s runtime system first creates various versions of proof methods specified by the strategy,
each of which tailored out for the proof obligation,
and combine them both sequentially and non-deterministically,
while exploring search space by applying these created proof methods.

The default strategy, \texttt{try\_hard}, outperformed, \texttt{sledgehammer},
the state-of-the-art proof automation for Isabelle/HOL,
by 16 percentage points when tested against 1,526 proof obligations for 300 seconds of timeout;
However, the dependence on the fixed default strategy
impairs \psl{}'s runtime system:
%causes the sub-optimal 
%behaviour of \psl{}'s runtime system:
\texttt{try\_hard} sometimes produces proof methods that are, for human engineers, 
obviously inappropriate to the given proof obligations.

%the application of the fixed default strategy to various sorts of proof obligations
%inevitably causes \psl{}'s runtime to wastes its search time 
%for inappropriate sub-strategies to those obligations.

\paragraph{\pamper{}}
Nagashima \etal{} developed \pamper{} \cite{pamper}, 
a \underline{p}roof \underline{m}ethod \underline{r}ecommendation tool,
trying to further automate proof development in Isabell/HOL.
\pamper{} learns when to use which proof methods 
from human-written large proof corpora called 
the Archive of Formal Proofs (AFP)\cite{AFP}.
The AFP is an online journal that hosts various formalization projects and 
mechanized proof scripts.
Currently, the AFP consists of 460 articles with 126,100 lemmas 
written by 303 authors in total.

\pamper{} first preprocess this data base:
it applies 108 assertions to each (possibly intermediate) proof obligation 
appearing in the AFP and
converts each of them into a vector of boolean values.
This way, \pamper{} creates 425,334 data points,
each of which is tagged with the name of proof method 
chosen by a human engineer to attack the obligation represented by the corresponding vector.
Then, \pamper{} applies a multi-output regression tree construction algorithm
to the database.
This process builds a regression tree for each proof method.
For instance, \pamper{} builds the following tree for the \induct{} method:
\begin{verbatim} 
(1, (10, expectation 0.0110944442872, 
         expectation 0.00345987448177),
    (10, expectation 0.0510162518838,
         expectation 0.0102138733024))
\end{verbatim}

where each of \verb|1| and \verb|10| in the first elements of the pairs represent the
number of the corresponding assertion.
For example, this tree tells that for proof obligations 
to which the assertion 1 returns false but the assertion 10 returns true,
the chance of an experienced proof engineer using the \induct{} method is about 5.1\%.

When a user of \pamper{} seeks for a recommendation, 
\pamper{} transforms the proof obligation at hand into a vector of boolean values
and looks up the trees and presents its recommendations.

\pamper{}'s regression tree construction is based on a problem transformation method, 
which handles a multi-output problem as a set of 
independent single-output problems:
For each obligation, 
\pamper{} attempts to provide multiple promising proof methods to attack the obligation,
by computing how likely each proof method is useful to the obligation one by one.

\paragraph{\pamper{} is not optimal to guide \psl{}}

One would imagine that it is natural step forward to improve \psl{}'s default strategy
by allowing \pamper{} to choose the most promising strategy for a given problem
instead of always applying the fixed strategy, \texttt{try\_hard}, naively.

Despite the positive results of cross-validation reported by Nagashima \etal{}, 
\pamper{}'s recommendation is not necessarily optimal to guide an automatic 
meta-tool based proof search for two reasons.
First, \pamper{} recommends only one step of proof method application,
even though many proof methods, such as \texttt{induction},
can discharge proof obligations only when followed by appropriate proof methods, such as \texttt{auto},
which is a general purpose proof method in Isabelle/HOL.
Second, when \pamper{} transforms a multi-output problem to a set of single-output problems,
\pamper{} preprocess the database introducing 
a conservative estimate of the correct choice of proof methods.
In the above example, \pamper{}'s pre-processor produces the following data point 
for all databases corresponding to proof methods that are not \verb|induction|.
\begin{verbatim}
not, [1,0,0,1,0,0,0,0,1,0,0,1,0,...]    
\end{verbatim}
%On one hand, we know that t
We know that this conservative estimate wrongfully lowers the expectation
for other proof methods for this case.
For example, Isabelle/HOL has multiple proof methods for induction,
such as \texttt{induct} and \texttt{induct\_tac}.
Experienced engineers know 
\texttt{induction} is a valid choice for most proof obligations where
\texttt{induct} is used.
Unfortunately, it is not computationally plausible to 
find out all alternative proofs for a proof obligation,
since many proof methods return intermediate proof obligations that have to be discharged by
other methods and even equivalently effective methods for the same obligation
may return distinct intermediate proof obligations.
In the above example, even though both \texttt{induct} and \texttt{induction} are
the right choice for many proof obligations, they return slightly different intermediate 
proof goals for most of the cases,
making it difficult to decide systematically if
\induct{} was also the right method where human engineers used \induction{} method.

\section{Evolutionary Prover in Isabelle/HOL}
We propose a novel approach based on evolutionary computation 
to overcome the aforementioned limitations of method recommendation 
based on supervised learning.
Our objective is to discover heuristics to choose
the most promising \psl{} strategy out of many hand written default strategies
when applied to a given proof goal,
so that \psl{} can exploit computational resources more effectively.

We represent programs as a sequence of floating point numbers,
each of which corresponds to a combinations of results of applying assertions to a proof obligation.
\pamper{} leaned 239 proof methods from the AFP and built a tree of height of two for each of them;
Therefore, we represent a program as a sequence of floating number of length 956,
which is the total number of leaf nodes in all regression trees.
Then, we assign such sequence to each default proof strategy.
Our prover first applies assertions to categorize a proof obligation,
then determines and applies the most promising strategy for that obligation.

As a training data set, we randomly picks up a set of proof obligations from large proof corpora.
And we measure how many obligations in this data set each version of our prover can discharge
given a fixed timeout for each obligation.
The more proof goals in the data set a prover can discharge, the better the prover is.

After each iteration, we mutate the program, which is a mapping function 
from a combination of results of assertions 
to the likelihood of each strategy being promising to the corresponding proof obligations.
After each evaluation, we select provers with higher success rates and
leave them for the next iteration, while discarding those with lower success rates.

We are still %at an early stage of this project and 
designing the details of the aforementioned experiment.
%The aforementioned experiment requires large computational resources 
%than most conventional theorem proving projects.
%Even though we plan to reuse the existing framework %of \psl{} and \pamper{}
%implemented for Isabelle/HOL,
%our approach should be transferable to other ITPs.
We expect that
when combined with the goal-oriented conjecturing mechanism \cite{pgt}
this project leads to the meta-tool based smart proof search in Isabelle/HOL initially proposed in 2017 \cite{smart_proof}.
%We expect that this project contributes to the realization of
%the meta-tool based smart proof search in Isabelle/HOL 
%.

%\end{document}  % This is where a 'short' article might terminate

%\begin{acks}
%  The authors would like to thank Dr. Yuhua Li for providing the
%  MATLAB code of the \textit{BEPS} method.
%
%  The authors would also like to thank the anonymous referees for
%  their valuable comments and helpful suggestions. The work is
%  supported by the \grantsponsor{GS501100001809}{National Natural
%    Science Foundation of
%    China}{http://dx.doi.org/10.13039/501100001809} under Grant
%  No.:~\grantnum{GS501100001809}{61273304}
%  and~\grantnum[http://www.nnsf.cn/youngscientists]{GS501100001809}{Young
%    Scientists' Support Program}.
%
%\end{acks}

\bibliographystyle{ACM-Reference-Format}
\bibliography{sample-bibliography} 


\begin{thebibliography}{11}

%%% the Web address.

\ifx \showCODEN    \undefined \def \showCODEN     #1{\unskip}     \fi
\ifx \showDOI      \undefined \def \showDOI       #1{#1}\fi
\ifx \showISBNx    \undefined \def \showISBNx     #1{\unskip}     \fi
\ifx \showISBNxiii \undefined \def \showISBNxiii  #1{\unskip}     \fi
\ifx \showISSN     \undefined \def \showISSN      #1{\unskip}     \fi
\ifx \showLCCN     \undefined \def \showLCCN      #1{\unskip}     \fi
\ifx \shownote     \undefined \def \shownote      #1{#1}          \fi
\ifx \showarticletitle \undefined \def \showarticletitle #1{#1}   \fi
\ifx \showURL      \undefined \def \showURL       {\relax}        \fi
% invisible to TeX
\providecommand\bibfield[2]{#2}
\providecommand\bibinfo[2]{#2}
\providecommand\natexlab[1]{#1}
\providecommand\showeprint[2][]{arXiv:#2}

\bibitem[\protect\citeauthoryear{Gonthier}{Gonthier}{2007}]%
        {4colour}
\bibfield{author}{\bibinfo{person}{Georges Gonthier}.}
  \bibinfo{year}{2007}\natexlab{}.
\newblock \showarticletitle{The Four Colour Theorem: Engineering of a Formal
  Proof}. In \bibinfo{booktitle}{{\em Computer Mathematics, 8th Asian
  Symposium, {ASCM} 2007, Singapore, December 15-17, 2007. Revised and Invited
  Papers}} {\em (\bibinfo{series}{Lecture Notes in Computer Science})},
  \bibfield{editor}{\bibinfo{person}{Deepak Kapur}} (Ed.),
  Vol.~\bibinfo{volume}{5081}. \bibinfo{publisher}{Springer},
  \bibinfo{address}{Berlin, Heidelberg}, \bibinfo{pages}{333}.
\newblock
\showDOI{%
\url{https://doi.org/10.1007/978-3-540-87827-8\_28}}


\bibitem[\protect\citeauthoryear{Hales, Adams, Bauer, Dang, Harrison, Hoang,
  Kaliszyk, Magron, McLaughlin, Nguyen, Nguyen, Nipkow, Obua, Pleso, Rute,
  Solovyev, Ta, Tran, Trieu, Urban, Vu, and Zumkeller}{Hales
  et~al\mbox{.}}{2015}]%
        {kepler}
\bibfield{author}{\bibinfo{person}{Thomas~C. Hales}, \bibinfo{person}{Mark
  Adams}, \bibinfo{person}{Gertrud Bauer}, \bibinfo{person}{Dat~Tat Dang},
  \bibinfo{person}{John Harrison}, \bibinfo{person}{Truong~Le Hoang},
  \bibinfo{person}{Cezary Kaliszyk}, \bibinfo{person}{Victor Magron},
  \bibinfo{person}{Sean McLaughlin}, \bibinfo{person}{Thang~Tat Nguyen},
  \bibinfo{person}{Truong~Quang Nguyen}, \bibinfo{person}{Tobias Nipkow},
  \bibinfo{person}{Steven Obua}, \bibinfo{person}{Joseph Pleso},
  \bibinfo{person}{Jason~M. Rute}, \bibinfo{person}{Alexey Solovyev},
  \bibinfo{person}{An~Hoai~Thi Ta}, \bibinfo{person}{Trung~Nam Tran},
  \bibinfo{person}{Diep~Thi Trieu}, \bibinfo{person}{Josef Urban},
  \bibinfo{person}{Ky~Khac Vu}, {and} \bibinfo{person}{Roland Zumkeller}.}
  \bibinfo{year}{2015}\natexlab{}.
\newblock \showarticletitle{A formal proof of the Kepler conjecture}.
\newblock \bibinfo{journal}{{\em CoRR\/}}  \bibinfo{volume}{abs/1501.02155}
  (\bibinfo{year}{2015}).
\newblock
\showeprint[arxiv]{1501.02155}
\showURL{%
\url{http://arxiv.org/abs/1501.02155}}


\bibitem[\protect\citeauthoryear{Klein, Andronick, Elphinstone, Heiser, Cock,
  Derrin, Elkaduwe, Engelhardt, Kolanski, Norrish, Sewell, Tuch, and
  Winwood}{Klein et~al\mbox{.}}{2010}]%
        {sel4}
\bibfield{author}{\bibinfo{person}{Gerwin Klein}, \bibinfo{person}{June
  Andronick}, \bibinfo{person}{Kevin Elphinstone}, \bibinfo{person}{Gernot
  Heiser}, \bibinfo{person}{David Cock}, \bibinfo{person}{Philip Derrin},
  \bibinfo{person}{Dhammika Elkaduwe}, \bibinfo{person}{Kai Engelhardt},
  \bibinfo{person}{Rafal Kolanski}, \bibinfo{person}{Michael Norrish},
  \bibinfo{person}{Thomas Sewell}, \bibinfo{person}{Harvey Tuch}, {and}
  \bibinfo{person}{Simon Winwood}.} \bibinfo{year}{2010}\natexlab{}.
\newblock \showarticletitle{seL4: formal verification of an operating-system
  kernel}.
\newblock \bibinfo{journal}{{\em Commun. {ACM}\/}} \bibinfo{volume}{53},
  \bibinfo{number}{6} (\bibinfo{year}{2010}), \bibinfo{pages}{107--115}.
\newblock
\showDOI{%
\url{https://doi.org/10.1145/1743546.1743574}}


\bibitem[\protect\citeauthoryear{Klein, Nipkow, Paulson, and Thiemann}{Klein
  et~al\mbox{.}}{2004}]%
        {AFP}
\bibfield{author}{\bibinfo{person}{Gerwin Klein}, \bibinfo{person}{Tobias
  Nipkow}, \bibinfo{person}{Larry Paulson}, {and} \bibinfo{person}{Rene
  Thiemann}.} \bibinfo{year}{2004}\natexlab{}.
\newblock \bibinfo{booktitle}{}.
\newblock
\showISSN{2150-914x}
\showURL{%
\url{https://www.isa-afp.org/}}


\bibitem[\protect\citeauthoryear{Kumar, Myreen, Norrish, and Owens}{Kumar
  et~al\mbox{.}}{2014}]%
        {cakeml}
\bibfield{author}{\bibinfo{person}{Ramana Kumar}, \bibinfo{person}{Magnus~O.
  Myreen}, \bibinfo{person}{Michael Norrish}, {and} \bibinfo{person}{Scott
  Owens}.} \bibinfo{year}{2014}\natexlab{}.
\newblock \showarticletitle{CakeML: a verified implementation of {ML}}. In
  \bibinfo{booktitle}{{\em The 41st Annual {ACM} {SIGPLAN-SIGACT} Symposium on
  Principles of Programming Languages, {POPL} '14, San Diego, CA, USA, January
  20-21, 2014}}, \bibfield{editor}{\bibinfo{person}{Suresh Jagannathan} {and}
  \bibinfo{person}{Peter Sewell}} (Eds.). \bibinfo{publisher}{{ACM}},
  \bibinfo{address}{New York, NY, USA}, \bibinfo{pages}{179--192}.
\newblock
\showDOI{%
\url{https://doi.org/10.1145/2535838.2535841}}


\bibitem[\protect\citeauthoryear{Leroy}{Leroy}{2009}]%
        {compcert}
\bibfield{author}{\bibinfo{person}{Xavier Leroy}.}
  \bibinfo{year}{2009}\natexlab{}.
\newblock \showarticletitle{Formal verification of a realistic compiler}.
\newblock \bibinfo{journal}{{\em Commun. {ACM}\/}} \bibinfo{volume}{52},
  \bibinfo{number}{7} (\bibinfo{year}{2009}), \bibinfo{pages}{107--115}.
\newblock
\showDOI{%
\url{https://doi.org/10.1145/1538788.1538814}}


\bibitem[\protect\citeauthoryear{Nagashima}{Nagashima}{2017}]%
        {smart_proof}
\bibfield{author}{\bibinfo{person}{Yutaka Nagashima}.}
  \bibinfo{year}{2017}\natexlab{}.
\newblock \showarticletitle{Towards Smart Proof Search for Isabelle}.
\newblock \bibinfo{journal}{{\em CoRR\/}}  \bibinfo{volume}{abs/1701.03037}
  (\bibinfo{year}{2017}).
\newblock
\showeprint[arxiv]{1701.03037}
\showURL{%
\url{http://arxiv.org/abs/1701.03037}}


\bibitem[\protect\citeauthoryear{Nagashima and He}{Nagashima and He}{2018}]%
        {pamper}
\bibfield{author}{\bibinfo{person}{Yutaka Nagashima} {and}
  \bibinfo{person}{Yilun He}.} \bibinfo{year}{2018}\natexlab{}.
\newblock \showarticletitle{Pa{M}pe{R}: Proof Method Recommendation System for
  {I}sabelle/{HOL}}. In \bibinfo{booktitle}{{\em Proceedings of the 33rd
  ACM/IEEE International Conference on Automated Software Engineering}} {\em
  (\bibinfo{series}{ASE 2018})}. \bibinfo{publisher}{ACM},
  \bibinfo{address}{New York, NY, USA}, \bibinfo{pages}{362--372}.
\newblock
\showISBNx{978-1-4503-5937-5}
\showDOI{%
\url{https://doi.org/10.1145/3238147.3238210}}


\bibitem[\protect\citeauthoryear{Nagashima and Kumar}{Nagashima and
  Kumar}{2017}]%
        {psl}
\bibfield{author}{\bibinfo{person}{Yutaka Nagashima} {and}
  \bibinfo{person}{Ramana Kumar}.} \bibinfo{year}{2017}\natexlab{}.
\newblock \showarticletitle{A Proof Strategy Language and Proof Script
  Generation for Isabelle/HOL}. In \bibinfo{booktitle}{{\em Automated Deduction
  - {CADE} 26 - 26th International Conference on Automated Deduction,
  Gothenburg, Sweden, August 6-11, 2017, Proceedings}} {\em
  (\bibinfo{series}{Lecture Notes in Computer Science})},
  \bibfield{editor}{\bibinfo{person}{Leonardo de~Moura}} (Ed.),
  Vol.~\bibinfo{volume}{10395}. \bibinfo{publisher}{Springer},
  \bibinfo{address}{Cham}, \bibinfo{pages}{528--545}.
\newblock
\showDOI{%
\url{https://doi.org/10.1007/978-3-319-63046-5\_32}}


\bibitem[\protect\citeauthoryear{Nagashima and Parsert}{Nagashima and
  Parsert}{2018}]%
        {pgt}
\bibfield{author}{\bibinfo{person}{Yutaka Nagashima} {and}
  \bibinfo{person}{Julian Parsert}.} \bibinfo{year}{2018}\natexlab{}.
\newblock \showarticletitle{Goal-Oriented Conjecturing for Isabelle/HOL}. In
  \bibinfo{booktitle}{{\em Intelligent Computer Mathematics - 11th
  International Conference, {CICM} 2018, Hagenberg, Austria, August 13-17,
  2018, Proceedings}} {\em (\bibinfo{series}{Lecture Notes in Computer
  Science})}, \bibfield{editor}{\bibinfo{person}{Florian Rabe},
  \bibinfo{person}{William~M. Farmer}, \bibinfo{person}{Grant~O. Passmore},
  {and} \bibinfo{person}{Abdou Youssef}} (Eds.), Vol.~\bibinfo{volume}{11006}.
  \bibinfo{publisher}{Springer}, \bibinfo{pages}{225--231}.
\newblock
\showISBNx{978-3-319-96811-7}
\showDOI{%
\url{https://doi.org/10.1007/978-3-319-96812-4\_19}}


\bibitem[\protect\citeauthoryear{Paulson}{Paulson}{2015}]%
        {incomplete}
\bibfield{author}{\bibinfo{person}{Lawrence~C. Paulson}.}
  \bibinfo{year}{2015}\natexlab{}.
\newblock \showarticletitle{A Mechanised Proof of G{\"{o}}del's Incompleteness
  Theorems Using Nominal Isabelle}.
\newblock \bibinfo{journal}{{\em J. Autom. Reasoning\/}} \bibinfo{volume}{55},
  \bibinfo{number}{1} (\bibinfo{year}{2015}), \bibinfo{pages}{1--37}.
\newblock
\showDOI{%
\url{https://doi.org/10.1007/s10817-015-9322-8}}


\end{thebibliography}

\end{document}